\documentclass[sigconf,authorversion,nonacm]{acmart}
\AtBeginDocument{%
  }

\acmConference[]{}{}{}

\usepackage{hyperref}
\hypersetup{colorlinks=true,linkcolor=blue, linktocpage}

\begin{document}

\title[Retrieval-Augmented Search for Map Collections]{Retrieval-Augmented Search for \\ Large-Scale Map Collections with ColPali}

\author{Jamie Mahowald}
\email{mahowald.jamie@gmail.com}
\orcid{0009-0007-1731-497X}
\affiliation{%
  \institution{Personal}
  \country{USA}
}

\author{Benjamin Charles Germain Lee}
\email{bcgl@uw.edu}
\orcid{0000-0002-1677-6386}
\affiliation{%
  \institution{Information School, University of Washington}
  \city{Seattle}
  \state{Washington}
  \country{USA}}

\renewcommand{\shortauthors}{Mahowald and Lee}

\begin{abstract}
Multimodal approaches have shown great promise for searching and navigating digital collections held by libraries, archives, and museums. In this paper, we introduce map-ras: a retrieval-augmented search system for historic maps. In addition to introducing our framework, we detail our publicly-hosted demo for searching 101,233 map images held by the Library of Congress. With our system, users can multimodally query the map collection via ColPali, summarize search results using Llama 3.2, and upload their own collections to perform inter-collection search. We articulate potential use cases for archivists, curators, and end-users, as well as future work with our system in both machine learning and the digital humanities. Our demo can be viewed at: {\color{blue}{\url{http://www.mapras.com}}}.
\end{abstract}


\begin{teaserfigure}
  \includegraphics[width=\textwidth]{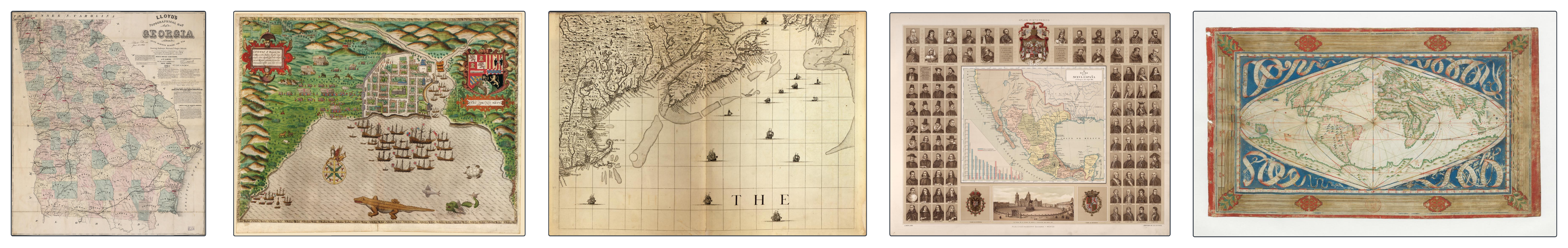}
  \caption{Historical maps from the Library of Congress's Geography \& Maps Division.}
  \label{fig:teaser}
\end{teaserfigure}


\maketitle

\section{Introduction}

Maps are key items for institutions that collect and preserve cultural heritage \cite{hessler}. They reveal historically significant insights into the priorities, capabilities, and limitations of the societies that created them, and so they serve as important subjects of  research. The Library of Congress (LOC), the world's largest map collector, has over 5.5 million maps \cite{zhang_2012}. Just over 10\% are digitized, while a quarter were received before machine-readable cataloging \cite{locResearchGuides}. 

Searching the LOC's Geography and Map (G\&M) digitized collection amounts to searching through textual metadata or text extracted by optical character recognition (OCR). This method requires thorough and accurate annotation of each searchable document. Moreover, it misses features specific to visual data not captured in the metadata, and it is fixed to the pre-defined hierarchy of map items. Maps are particularly difficult to characterize because they combine visual and textual elements, posing challenges for traditional, unimodal machine learning solutions.

Recent advances in multimodal machine learning have made this problem more tractable. The contrastive language-image pretraining (CLIP) model class \cite{radford_CLIP} in particular allows a user to compare a text prompt with a pre-computed corpus of image embeddings. By returning the images whose embeddings are most geometrically similar to the prompt embedding (i.e., nearest neighbors search), CLIP can function as a search engine. However, the base CLIP model struggles with text recognition \cite{clip_text_context}, and its hard labeling resists easy fine-tuning on similar-looking training data. 

In this work, we apply the ColPali document-retrieval framework \cite{colpali} to the task of searching maps. Broadly designed as an aid in retrieval-augmented generation (RAG) settings, ColPali has several properties that facilitate this task. Its soft similarity scores can better handle visually similar items, and its multi-vector representations can capture more complex relationships in data. As opposed to models whose embeddings capture only global features, ColPali can attend to specific regions of a document. This framework also makes it easy for users to dynamically upload and add to their corpus, a capability we call retrieval-augmented search (RAS).

In summary, we provide the following contributions:

\begin{enumerate}
    \item We introduce retrieval-augmented search, a variant of RAG, for cultural heritage collections, and we present a demo for 101,233 map images held by the Library of Congress. Our demo can be found at: {\color{blue}{\url{http://www.mapras.com}}}.
    \item We elaborate on use cases of our map retrieval-augmented search system and detail future directions for our research.
    \item We release our code at {\color{blue}{\url{https://github.com/j-mahowald/mapras}}}.
\end{enumerate}

\section{Related Work}

\subsection{Library of Congress Maps}

As our use case, we use the LOC G\&M's digital collection of 563,698 unique ``segments'' (unique images) divided among 57,962 ``resources'' (composite items). A small sampling can be found in Figure \ref{fig:teaser}. Items contain anywhere from one to 11,981 segments (though most are small, with median 2 segments per item), and metadata is recorded at the item level. These maps span centuries, as well as domains, from geography and topography to historical analysis. 

Of the 563,698 images, 439,947 (78\% of segments), belong to the collection of Sanborn fire insurance maps, which provide granular data on urban building infrastructure and land use \cite{sanborns}. Because these are easily indexable by metadata, we restrict our tool to the approximately 100,000 non-Sanborn maps in the Library's digital collection that return valid calls.

The G\&M online catalog is broadly searchable via metadata facets including collection, keyword, location, and time period. Each individual catalog entry can be viewed for each map, along with information about the creation and context.

\begin{figure}[!h]
\includegraphics[width=0.44\linewidth]{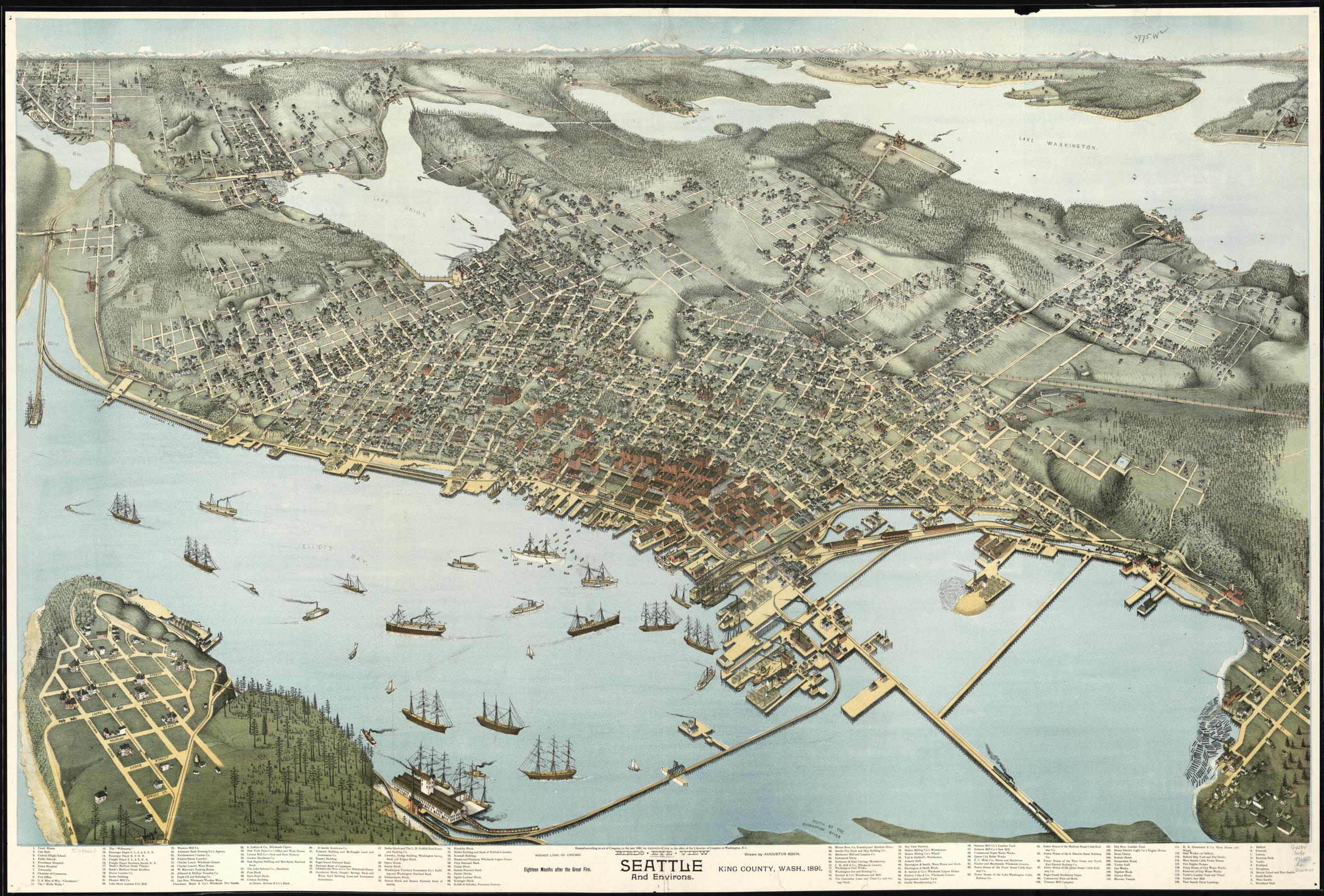}
\includegraphics[width=0.49\linewidth]{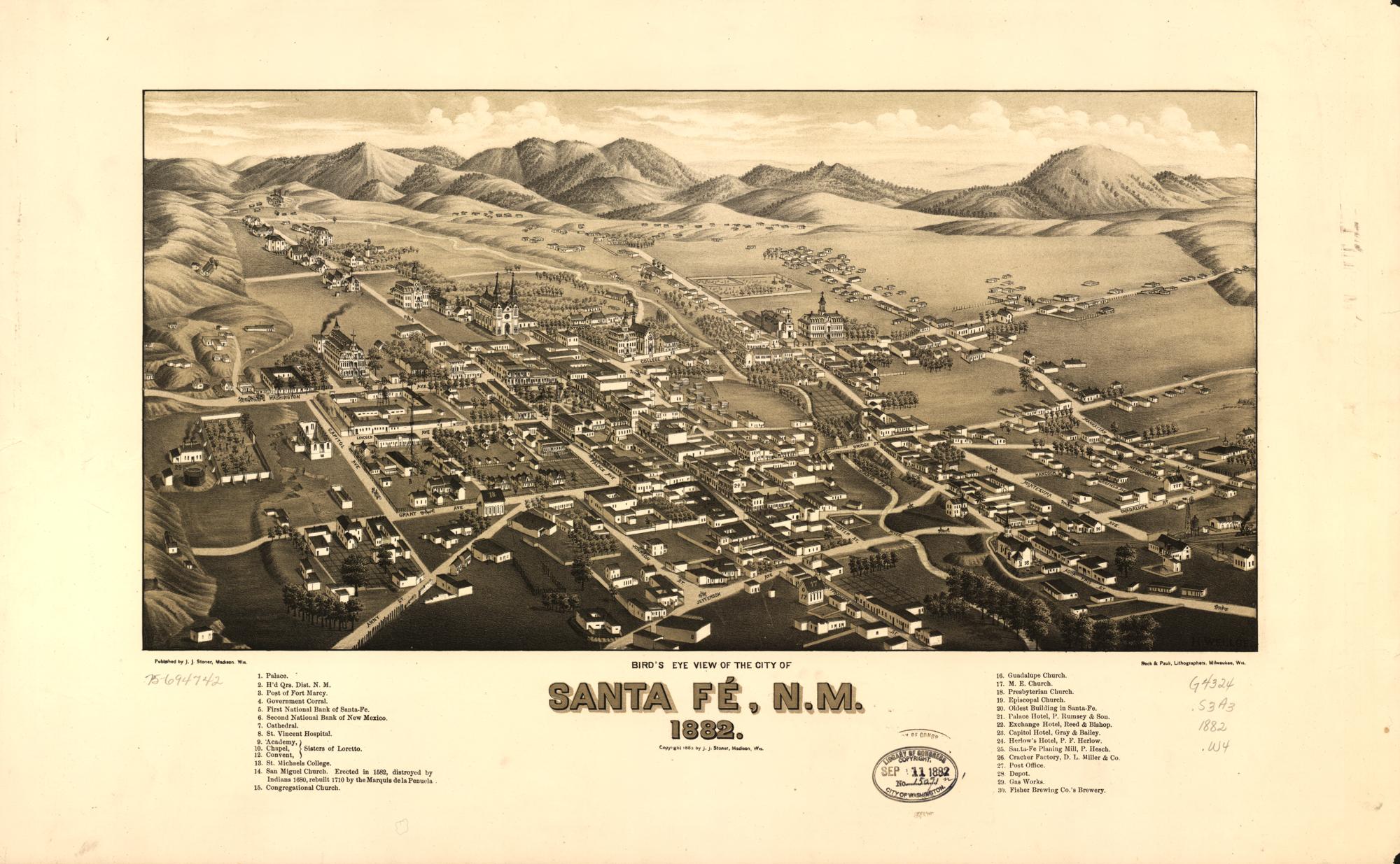}
\caption{Historical panoramic maps of Seattle, WA (1891), and Santa Fe, NM (1882), showing landmarks, street layouts, and historical buildings.}
\end{figure}

\subsection{ColPali}

ColPali offers two base models: the original ColPali built on Google's PaliGemma (a 3-billion-parameter vision-language model) \cite{paligemma}, and ColQwen built on Alibaba's Qwen2 VLM (2 billion parameters) \cite{colpali}. Both models use multiple patch embeddings to capture details across different sections of a page. ColQwen generates 768 patch embeddings per page, while ColPali generates 1024, though both use the same embedding dimension of 128. We use ColQwen in this implementation to reduce comptutational costs.

\subsection{Multimodality and Digital Humanities}

While map collections are routinely utilized by researchers across disciplines, searching maps oftentimes suffer from fundamental limitations \cite{topoographic_metadata}. For example, map metadata is never completely descriptive, and neither is the OCR-able text appearing on a given map. Given the sheer scale of digitized maps available, developing new ways to search these maps is more important than ever.

This paper builds on work in the digital humanities that has demonstrated the value of multimodal search for digital collections held by libraries, archives, and museums. Deemed the  ``multimodal turn'' \cite{multimodal_turn}, models such as CLIP \cite{radford_CLIP} have shown great promise for increased discoverability for digital collections ranging from photojournalism collections \cite{digital_collections_explorer} to lantern slides \cite{smits_kestemont}. This paper extends our previous work surrounding applying CLIP to digitized maps for search \& discovery, informed by the perspectives of staff at the Library of Congress \cite{mahowald_lee}. However, as described earlier, CLIP struggles with text recognition, especially with longer passages. For this reason, we have adopted ColPali.

\begin{figure}[!h]
\includegraphics[width=\linewidth]{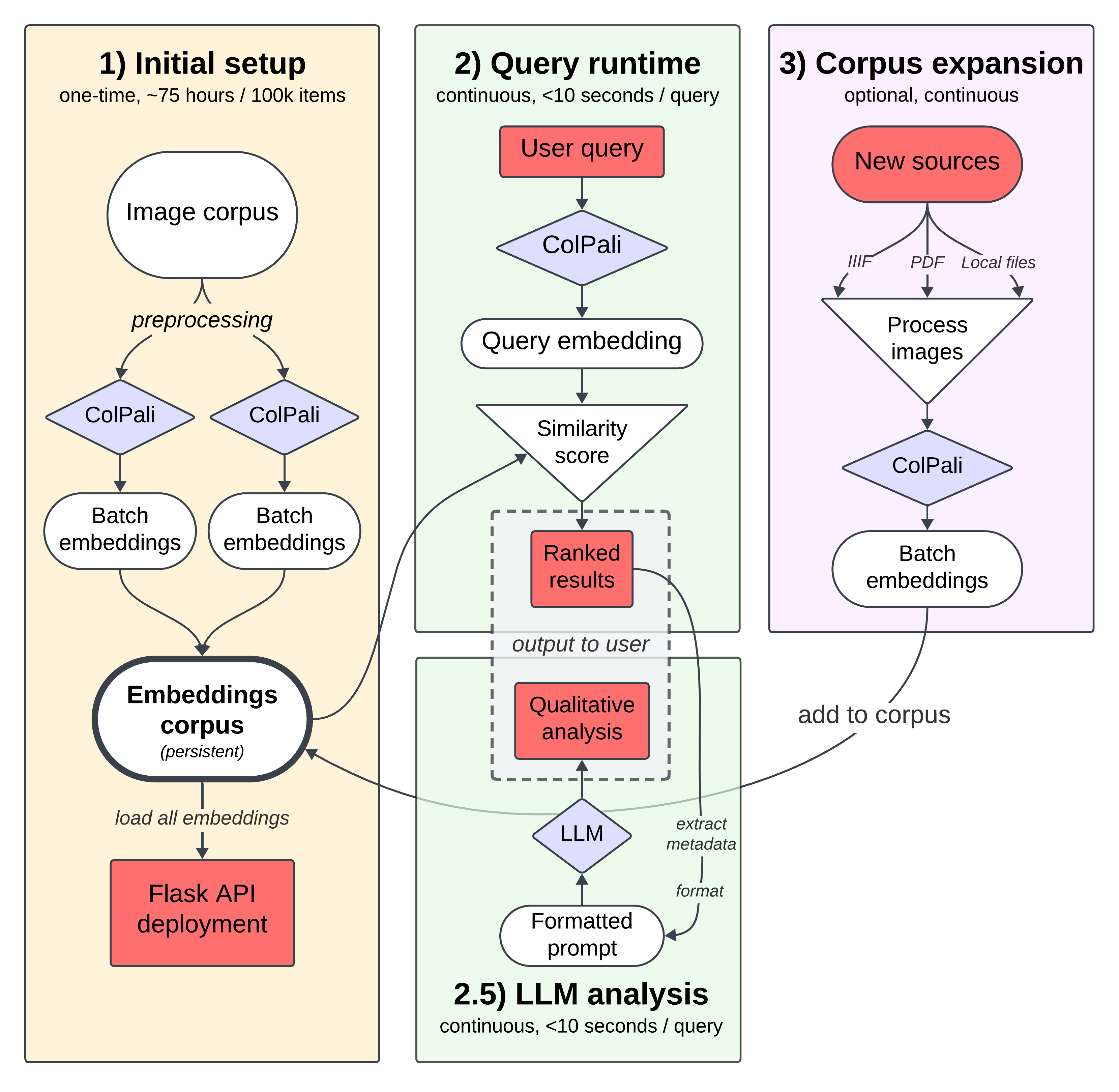}
\caption{A full flow chart of our 3-stage pipeline. Red shapes indicate direct interface with the user, while blue diamonds are models loaded onto the server. The embeddings corpus is the only persistent object at 28GB.}\label{fig:overview}
\end{figure}

A number of other works have begun exploring the application of ColPali to digital cultural heritage \cite{riverecho, fashionrna, artseek}. In addition to applying ColPali, we utilize Llama 3.2 \cite{llama3} to generate summaries of search results, in line with RAG-style search.

\section{Method}

Our fast and reusable retrieval-augmented search engine is motivated by the task of searching map data for cultural institutions: 
\begin{enumerate}
    \item First, to ensure accessibility for resource-constrained institutions, the tool should create high-quality embeddings of complex visual data quickly on consumer-grade hardware. The method should be available for practitioners who want to implement it on their own data.
    \item It should embed several modes into a common latent space to enable comparison between them.
    \item Because maps have both visual and textual data, it should distinguish between sections of a single document and parse them individually, rather than create a single global embedding.
    \item It should allow users to easily add new items to the corpus as they come up.
\end{enumerate}

As shown in Figure \ref{fig:overview}, our method follows three stages: data preparation, embedding generation, and search interface deployment. We describe all three stages below.

\begin{figure}[!h]
\includegraphics[width=\linewidth]{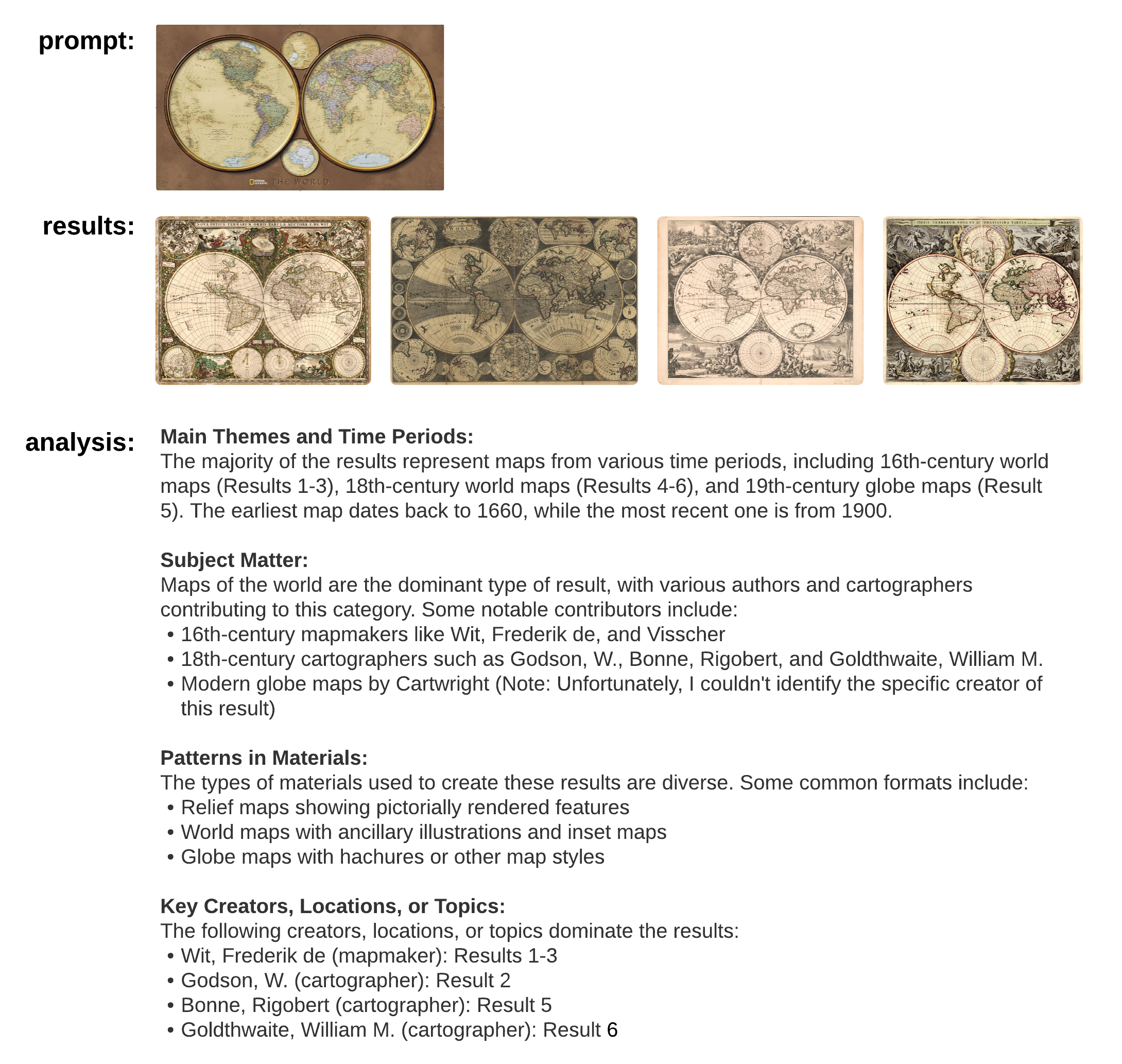}
\caption{Search on a non-LOC image results in visually similar images taken from the LOC's corpus. Analysis highlights details on the results taken from metadata.}
\label{fig:example_img}
\end{figure}
\begin{figure}[!h]
\includegraphics[width=\linewidth]{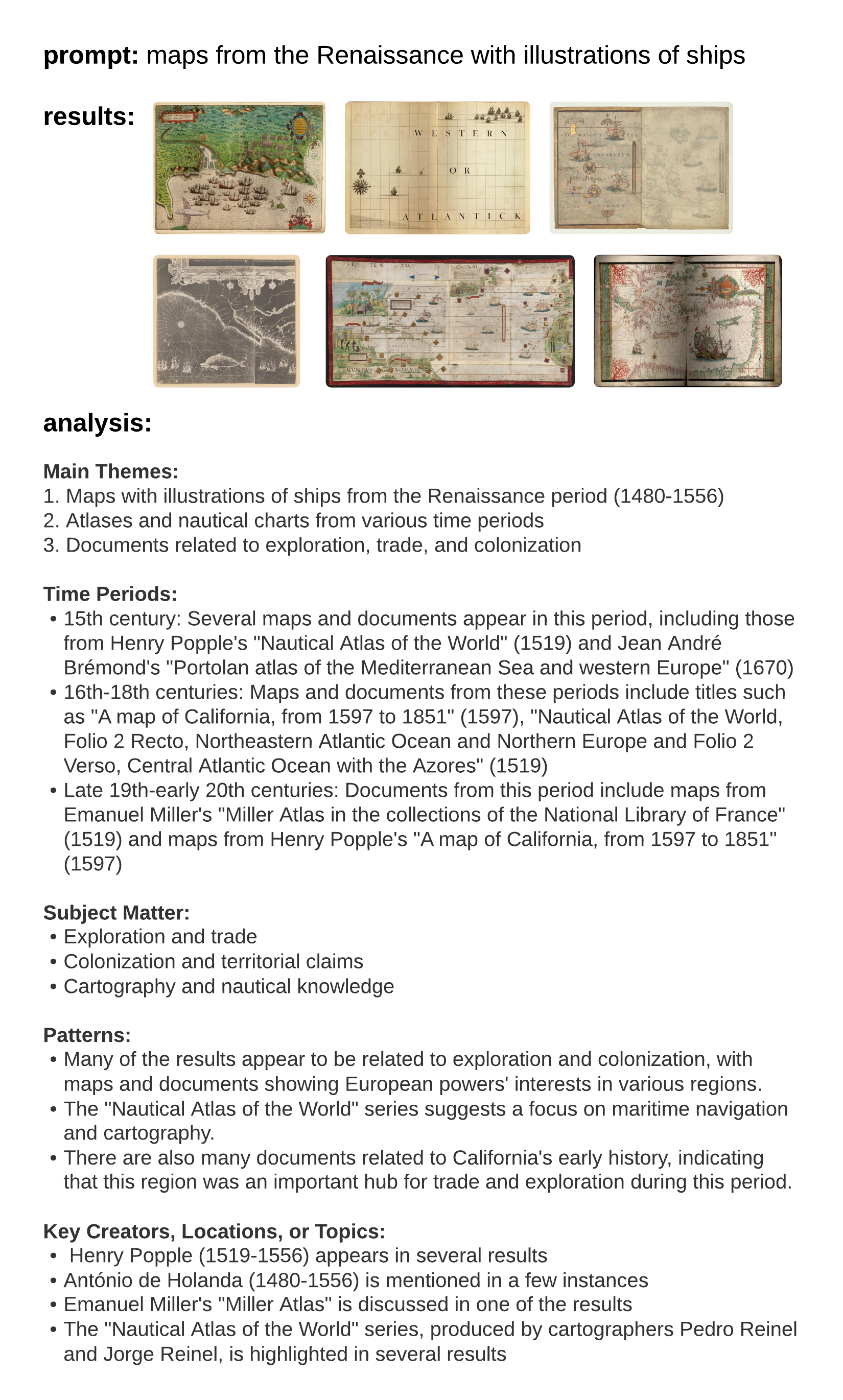}
\caption{The tool allows us to discern features on a map like illustrations of ships that do not appear in the items' metadata.}
\label{fig:example_text}
\end{figure}

\subsection{Data Preparation}

Working from a central CSV file, we download batches of 500 images sequentially, accumulating them in temporary storage and processing them in sub-batches to respect GPU memory constraints. In keeping with cultural heritage norms, we work in the International Image  Interoperability Framework (IIIF) throughout our pipeline. 

\subsection{Embedding Generation}

Working in batches, we embed each image in the corpus using the pre-trained ColQwen2 model. On a single NVIDIA T4 Tensor Core GPU, a single embedding processes in 3 to 4 seconds, with its model requiring 5.62GB of GPU memory.\footnote{On an AWS \texttt{g4dn.4xlarge} instance, this requires 75 hours of total compute, amounting to approximately $\$90$ dollars.} Each embedding is put into a dictionary with its unique IIIF identifier, which can be used to reference relevant metadata from the master CSV. Embeddings are saved at the batch level in distributed pickle files.

\subsection{Search Interface}

Once a base set of embeddings is processed, we deploy a Flask-based REST API that loads all distributed embeddings into memory on startup. The API exposes several endpoints for corpus management and querying, with the primary search endpoint accepting POST requests containing a natural language query string and a parameter $k$ specifying the number of results to return (default: 5).

Query processing follows the late-interaction paradigm: the user's text query is embedded using the same ColQwen2 model, then scored against all document embeddings using the processor's MaxSim computation, which determines the maximum similarity between each query token vector and all document patch vectors, preserving fine-grained semantic matching while remaining computationally tractable at the scale of the corpus.

In Figures \ref{fig:example_img} and \ref{fig:example_text}, we show a screenshot of an example search, along with several top results.
Results are ranked by similarity score and returned with metadata including the document title, the image itself, the `loc.gov` API resource, the document type, and numerical score. Response times for queries against our corpus of 25,000+ embedded images average under one second after the initial model load, making the system suitable for interactive exploration. 

On a server with a single NVIDIA T4 Tensor Core GPU for AI inference, submitting a query through the 120,000-item embedding corpus takes roughly 6 seconds. Indeed, inference time grows linearly with corpus size, demanding new techniques for small-scale inference once corpus size exceeds a few hundred thousand. We leave this for future work.

\subsubsection{Image search}

Because vision-language models embed visual and textual data to a common embedding space, the tool can easily accept input images for a reverse-image-search functionality. After images are processed, this follows the same scoring regime as text-to-image search. Notably, the model does not transfer between modes in this case, and the similarity scores between an image query and its result tend to be an order of magnitude higher than between a text query and its result. 

\subsubsection{Retrieval-augmented Search Framework}\label{sec:framework}
Our implementation follows a retrieval-augmented generation (RAG) approach that allows dynamic expansion of the corpus. Users can augment the base set with their own documents, whether from partner institutions' IIIF servers, local digitized materials, or PDF publications, during or between search sessions. These personalized search contexts allow a user to compare their own materials to those of the corpus, and to customize the search engine for their own uses.

More broadly, persistent corpus expansion allows institutions to collaboratively build federated search indices, where multiple organizations contribute embeddings from their collections without centralizing the underlying image files, respecting data sovereignty while enabling cross-institutional discovery.

All additions, regardless of origin, are processed through the identical ColQwen2 embedding pipeline, projected into the same 128-dimensional vector space, and scored using the same late-interaction mechanism. This ensures that user-contributed materials are semantically comparable to the base corpus.

\subsubsection{LLM Analysis}

Lastly, we include a feature that allows a user to generate qualitative LLM analysis on the results of their search. After a search is completed, the pipeline extracts metadata from the search result, formats it into a prompt, and passes it to a lightweight Llama 3.2-1B, which provides short, digestible information on the major themes, time periods, formats, subjects, and authors of the results. Analysis typically runs for 5-15 seconds, depending on the number of results requested. 

\section{Use Cases}

We envision our retrieval-augmented search system for large-scale map collections to have a number of uses for both collection curators and end-users. On the curatorial side, archivists, curators, catalogers and collection stewards responsible for digitizing collections and providing access to them typically rely on traditional manual methods of cataloging and producing finding aids. This process is time-intensive, especially for larger-scale collections, such as the ones held by the Library of Congress. Put simply, approaching the scale of half a million map images necessitates additional methods to draw connections across a digital collection. With our retrieval-augmented search system, archivists, curators, and catalogers can identify patterns at scale and also interact with the collection in new, multimodal ways. They can test hypotheses about the prevalence of different features during finding aid construction and also make associations that otherwise might not be possible by relying on existing metadata. As digitization efforts continue to grow, and born-digital collections approach orders of magnitude larger, such methods will only become more important.

Similarly, end-users from academic researchers to members of the public face the challenge of how to make sense of large-scale digital collections. Academics searching map collections for specific motifs, aesthetic qualities, or accompanying text might encounter the analogous problem with existing forms of search. Our system enables end-users to define more flexible queries, generate relevant summarizations, and more generally see collections at scales not possible with basic metadata search (i.e., ``distant viewing'' \cite{distant_viewing}).

Lastly, we highlight the possibilities for integrating \textit{inter-}collection search. As described in Section \ref{sec:framework}, our system enables others to dynamically expand the collection included. The possibilities with this dynamic corpus expansion functionality are manifold. First, archivists and collection stewards can utilize this approach to search \textit{across} collections that are ordinarily siloed. Second, digital humanities scholars can use this approach for exploratory analysis in order to identify similarities and differences between collections. Third, as articulated in Section \ref{sec:framework}, we see a broader vision of enabling the multi-institutional construction of federated search indices using these vectorized approaches.

\section{Future Work}

We conclude by offering a few directions for future work. First, we plan to expand our retrieval-augmented search system into a full vision language model-based RAG system by generating summaries based on what the model finds from the map images directly, not just the map metadata. Second, we will explore finetuning ColPali for specific use with digital cultural heritage collections. Third, we hope to expand our system into a more flexible system for searching a range of digital cultural heritage collections, as well as conduct user evaluations of it. We encourage users to visit {\color{blue}{\url{http://www.mapras.com}}} and test their own prompts, and  welcome feedback.

\bibliographystyle{ACM-Reference-Format}
\bibliography{bibliography}

\end{document}